\documentclass[sigconf,anonymous=false]{acmart}
\AtBeginDocument{%
  \providecommand\BibTeX{{%
    \normalfont B\kern-0.5em{\scshape i\kern-0.25em b}\kern-0.8em\TeX}}}

\setcopyright{acmcopyright}
\copyrightyear{2022}
\acmYear{2022}
\acmDOI{XXXXXXX.XXXXXXX}

\acmConference{DLP-KDD 2022}{August 14, 2022}{Washington DC, USA}
\acmBooktitle{Washington DC 2022: ACM SIGKDD Conference on Knowledge Discovery and Data Mining, August 14-18, 2022, Washington DC}
\acmPrice{15.00}
\acmISBN{978-1-4503-XXXX-X/18/06}

\usepackage{hyperref}
\usepackage{cleveref}
\usepackage{natbib}
\usepackage{amsmath}
\usepackage{enumitem}
\usepackage{subcaption}
\begin{document}

\title[Research Track Paper]{DynInt: Dynamic Interaction Modeling for Large-scale Click-Through Rate Prediction}

\author{Yachen Yan}
\authornote{Both authors contributed equally to this research.}
\email{yachen.yan@creditkarma.com}
\orcid{1234-5678-9012}
\author{Liubo Li}
\authornotemark[1]
\email{liubo.li@creditkarma.com}
\affiliation{%
    \institution{Credit Karma}
    \streetaddress{760 Market Street}
    \city{San Francisco}
    \state{California}
    \country{USA}
    \postcode{94012}
}

\renewcommand{\shortauthors}{Yachen and Liubo}
\renewcommand{\subtitle}{DynInt}

\begin{abstract}
Learning feature interactions is the key to success for the large-scale CTR prediction in Ads ranking and recommender systems. In industry, deep neural network-based models are widely adopted for modeling such problems. Researchers proposed various neural network architectures for searching and modeling the feature interactions in an end-to-end fashion. However, most methods only learn static feature interactions and have not fully leveraged deep CTR models' representation capacity. In this paper, we propose a new model: DynInt. By extending Polynomial-Interaction-Network (PIN), which learns higher-order interactions recursively to be dynamic and data-dependent, DynInt further derived two modes for modeling dynamic higher-order interactions: dynamic activation and dynamic parameter. In dynamic activation mode, we adaptively adjust the strength of learned interactions by instance-aware activation gating networks. In dynamic parameter mode, we re-parameterize the parameters by different formulations and dynamically generate the parameters by instance-aware parameter generation networks. Through instance-aware gating mechanism and dynamic parameter generation, we enable the PIN to model dynamic interaction for potential industry applications. We implement the proposed model and evaluate the model performance on real-world datasets. Extensive experiment results demonstrate the efficiency and effectiveness of DynInt over state-of-the-art models.
\end{abstract}




\maketitle

\section{Introduction}
Click-through rate (CTR) prediction model~\cite{richardson2007predicting} is an essential component for the large-scale search ranking, online advertising and recommendation system~\cite{mcmahan2013ad,he2014practical,cheng2016wide,zhang2019deep}.

Many deep learning-based models have been proposed for CTR prediction problems in the industry and have become dominant in learning the useful feature interactions of the mixed-type input in an end-to-end fashion\cite{zhang2019deep}. While most of the existing methods focus on automatically modeling static deep feature representations, there are very few efforts on modeling feature representations dynamically.

However, the static feature interaction learning methods may not fully fulfill the complexity and long-tail property of large-scale search ranking, online advertising, and recommendation system problems. The static feature interaction learning can not capture the characteristics of long-tail instances effectively and efficiently, as short-tail instances and high-frequency features will easily dominate the shared and static parameters of the model. Therefore, we argue that the learned feature interactions should be dynamic and adaptive to different instances for capturing the pattern of long-tail instances.

Inspired by the dynamic model framework, we propose a dynamic feature interaction learning method called DynInt. DynInt has two schemes that enhance the base model (xDeepInt~\cite{yan2020xdeepint}) to learn the dynamic and personalized interactions.
\begin{itemize}[leftmargin=10pt]
    \item DynInt-DA: We reinforce the Polynomial-Interaction-Network by instance-aware activation gating networks for adaptively refining learned interactions of different s.
    \item DynInt-DGP and DynInt-DWP: We re-parametrize the parameters of Polynomial-Interaction-Network by instance-aware parameter generation networks for adaptively generating and re-weighting the parameters to learn dynamic interactions.
    \item We introduce the different computational paradigms to reduce the memory cost and improve efficiency in implementations effectively.
    \item We introduce the orthogonal regularization for enhancing the diversity of learned representation in dynamic interaction modeling settings.
    \item We conduct extensive experiments on real-world datasets and study the effect of different hyper-parameters, including but not limited to: kernel size, latent rank, and orthogonal regularization rate.
\end{itemize}

\section{Related Work}

\subsection{Static Interaction Modeling}
Most deep CTR models map the high-dimensional sparse categorical features and continuous numerical features onto a low dimensional latent space as the initial step. Many existing model architectures focus on learning static implicit/explicit feature interactions simultaneously.

Various hybrid network architectures~\cite{cheng2016wide,qu2016product,qu2018product,wang2017deep,wang2021dcn,guo2017deepfm,lian2018xdeepfm} utilize feed-forward neural network with non-linear activation function as its deep component, to learn implicit interactions. The complement of the implicit interaction modeling improves the performance of the network that only models the explicit interactions~\cite{beutel2018latent}. However, this type of approach fails to bound the degree of the learned interactions.

Deep \& Cross Network (DCN)~\cite{wang2017deep} and its improved version DCN V2  ~\cite{wang2021dcn} explores the feature interactions at the bit-wise level explicitly in a recursive fashion. Deep Factorization Machine (DeepFM)~\cite{guo2017deepfm} utilizes factorization machine layer to model the pairwise vector-wise interactions. Product Neural Network (PNN)~\cite{qu2016product,qu2018product} introduces the inner product layer and the outer product layer to learn vector-wise interactions and bit-wise interactions, respectively. xDeepFM~\cite{lian2018xdeepfm} learns the explicit vector-wise interaction by using Compressed Interaction Network (CIN), which has an RNN-like architecture and learns vector-wise interactions using Hadamard product. FiBiNET~\cite{huang2019fibinet} utilizes Squeeze-and-Excitation network to dynamically learn the importance of features and model the feature interactions via bilinear function. AutoInt~\cite{song2018autoint} leverages the Transformer~\cite{vaswani2017attention} architecture to learn different orders of feature combinations of input features. xDeepInt~\cite{yan2020xdeepint} utilizes polynomial interaction layer to recursively learn higher-order vector-wise and bit-wise interactions jointly with controlled degree, dispensing with jointly-trained DNN and nonlinear activation functions.

\subsection{Gating Mechanism in Deep Learning}
Gating mechanism is widely used in computer vision and natural language processing, more importantly, CTR prediction for Ads ranking and recommendation systems.

LHUC~\cite{swietojanski2014learning} learns hidden unit contributions for speaker adaptation. Squeeze-and-Excitation Networks~\cite{hu2018squeeze} recalibrates channel-wise feature responses by explicitly modeling interdependencies and multiplying each channel with learned gating values. Gated linear unit (GLU)~\cite{dauphin2017language} was utilized to control the bandwidth of information flow in language modeling.

Multi-gate Mixture-of-Experts (MMoE)~\cite{ma2018modeling} utilizes gating networks to automatically weight the representation of shared experts for each task, so that simultaneously modeling shared information and modeling task specific information. GateNet~\cite{huang2020gatenet} leverages feature embedding gate and hidden gate to select latent information from the feature-level and intermediate representation level, respectively. MaskNet~\cite{wang2021masknet} introduces MaskBlock to apply the instance-guided mask on feature embedding and intermediate output. The MaskBlock can be combined in both serial and parallel fashion. Personalized Cold Start Modules (POSO) \cite{dai2021poso} introduce user-group-specialized sub-modules to reinforce the personalization of existing modules, for tackling the user cold start problem. More specifically, POSO combines the personalized gating mechanism with existing modules such as Multi-layer Perceptron, Multi-head Attention, and Multi-gated Mixture of Experts, and effectively improves their performance.

\subsection{Parameter Generation in Deep Learning}
Besides adjustments on feature embedding and intermediate representations, dynamic parameter generation is more straightforward in making the neural network adaptive to the input instance.

Dynamic filter networks (DFN)~\cite{jia2016dynamic} and HyperNetworks~\cite{ha2016hypernetworks} are two classic approaches for runtime parameter generation for CNNs and RNNs, respectively. DynamicConv~\cite{wu2019pay}, which is simpler and more efficient, was proved to perform competitively to the self-attention.

In the recommendation area, PTUPCDR~\cite{zhu2022personalized} leverages meta network fed with users’ characteristic embeddings to generate personalized bridge functions to achieve personalized transfer for cross domain recommendation. Co-Action Network (CAN)~\cite{bian2022can} uses learnable item embedding as parameters of micro-MLP to realize interaction with user historical behavior sequence, for approximating the explicit and independent pairwise feature interactions without introducing too many additional parameters. Adaptive Parameter Generation network (APG)~\cite{yan2022apg} re-parameterize the dense layer parameters of deep CTR models with low-rank approximation, where the specific parameters are dynamically generated on-the-fly based on different instances.

\section{Proposed Model: Dynamic Interaction Network (DynInt)}

\begin{figure}[htbp]
    \centering
    \includegraphics[width=0.25\textwidth, height=0.50\textwidth]{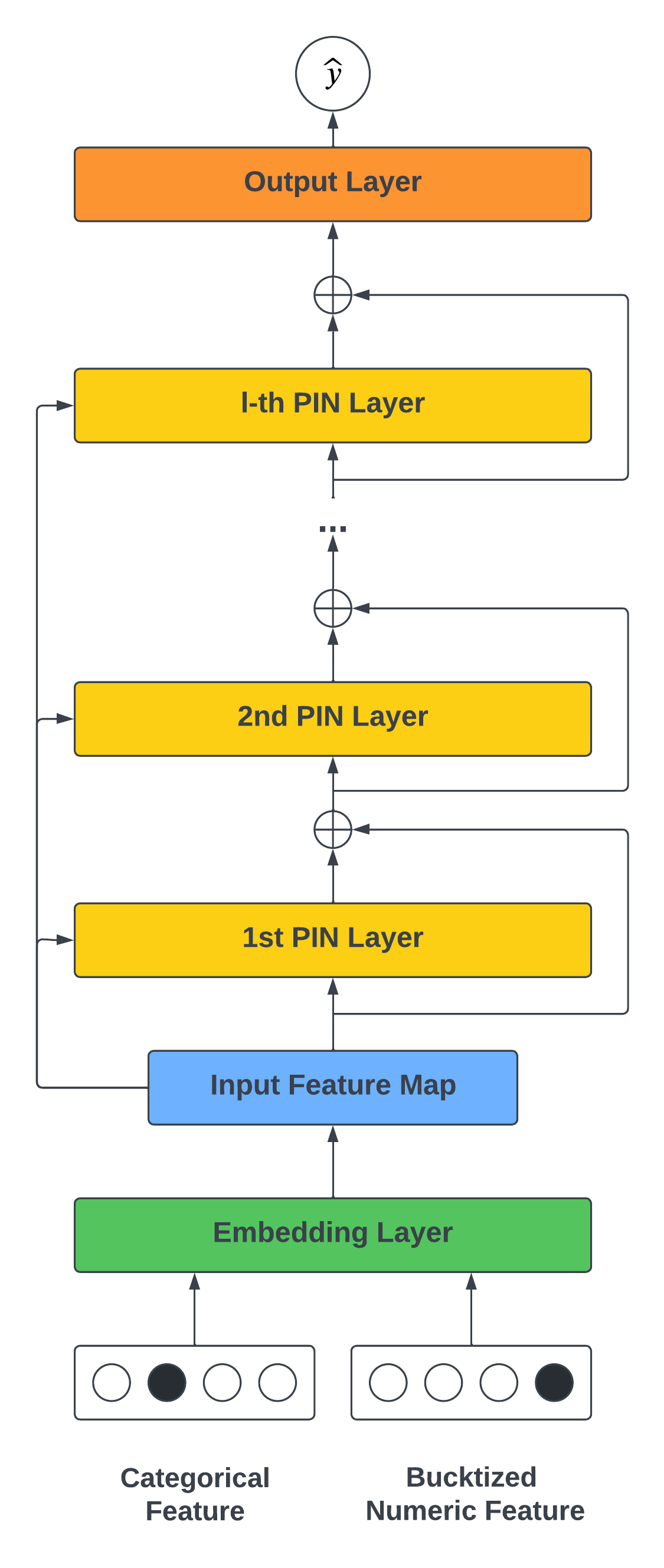}
    \caption{The architecture of unrolled Polynomial Interaction Network with residual connections}
    \label{fig_pin}
    \medskip
    \small
\end{figure}

In this section, we give an overview of the architectures of DynInt variants. We first clarify the notations. $F$ is the number of features. $i$ is the instance index. $j$ is the feature index. $D$ is the feature embedding size. $K$ is the rank of the approximation. $B$ is the batch size. $X_{0}$ is the input feature map and $X_l$ is the $l$-th layer output. $W_l$ is the $l$-th layer parameter matrix.

Our models first embed each feature into a dense embedding vector $x_{j}$ of dimension $D$. We denote the output of embedding layer $X_0 \in R^{F\times D}$ as the input feature map:
\begin{equation}
\begin{aligned}
	X_0 = [x_{0}, x_{1}, \cdots, x_{F-1}]^\intercal.
\end{aligned}
\end{equation}

\begin{figure}[htbp]
\centering
    \begin{subfigure}[b]{0.40\textwidth}  
    \centering
    \includegraphics[width=0.9\textwidth, height=0.25\textwidth]{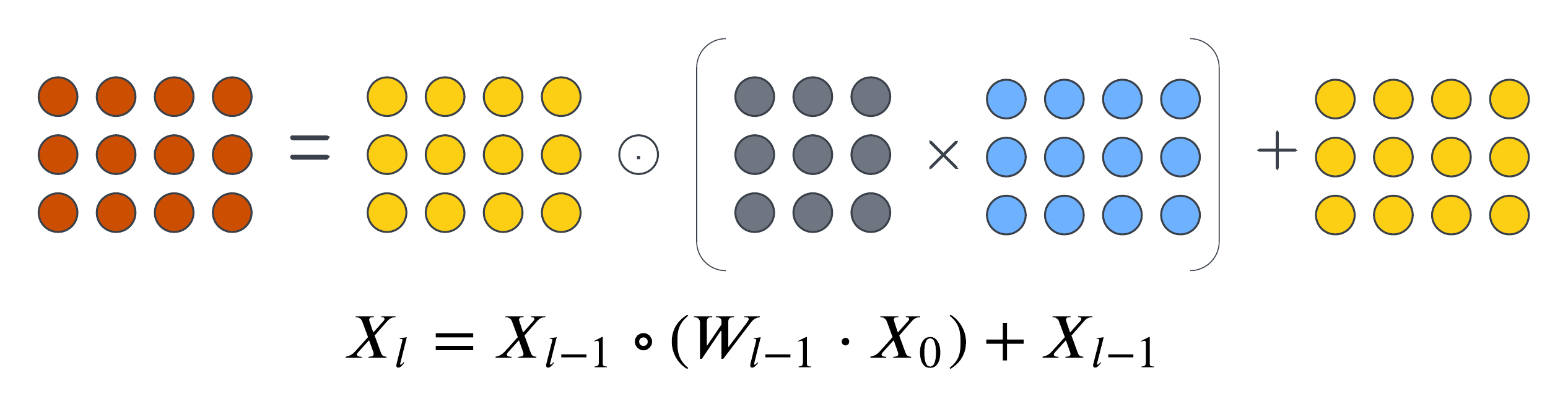}
    \caption{Polynomial Interaction Network (PIN) Layer}
    \label{fig_pin}
    \end{subfigure}
    \vfill
    \begin{subfigure}[b]{0.40\textwidth}  
    \centering
    \includegraphics[width=0.70\textwidth, height=0.30\textwidth]{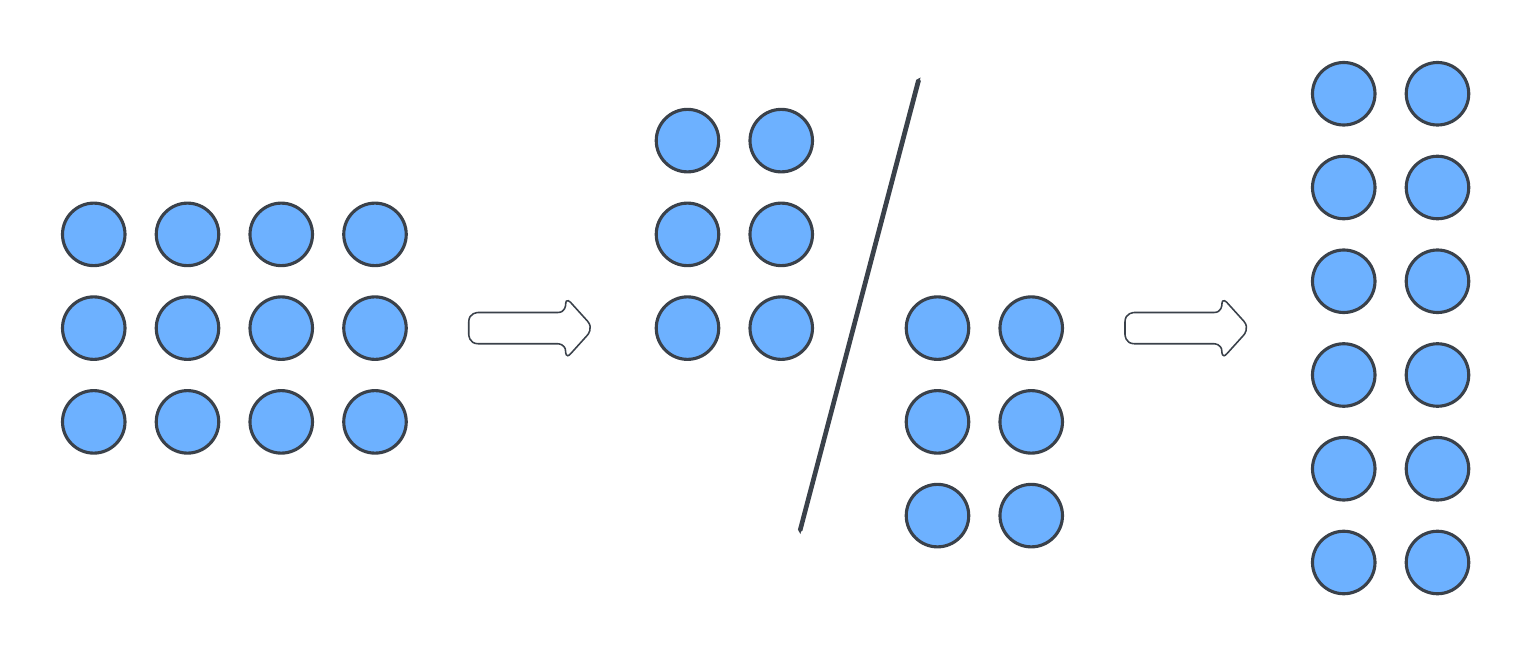}
    \caption{Subspace-crossing Mechanism}
    \label{fig_subspace}
    \end{subfigure}
    \vfill
    \begin{subfigure}[b]{0.35\textwidth}  
    \centering
    \includegraphics[width=0.90\textwidth, height=0.70\textwidth]{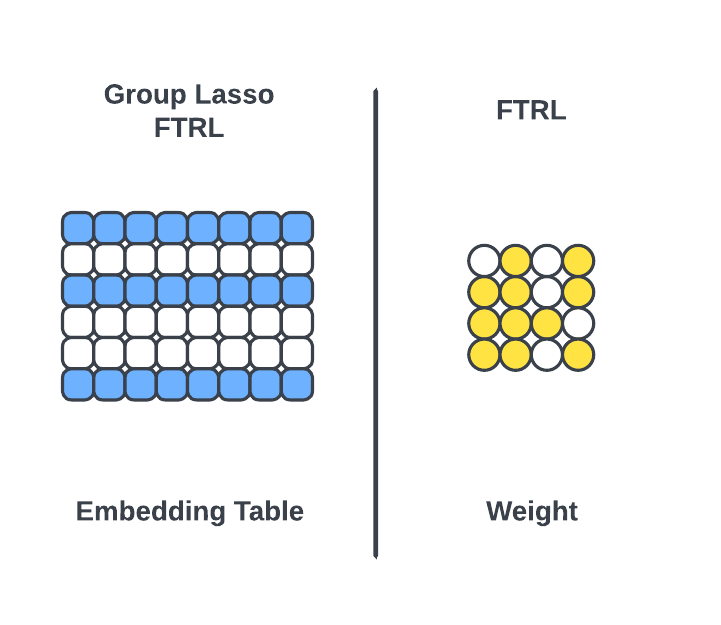}
    \caption{GroupLasso FTRL v.s. FTRL}
    \label{fig_ftrl}
    \end{subfigure}
\caption{The key components of the xDeepInt model}
\label{fig_pin_subspace_ftrl}
\end{figure}

The shared components of all the DynInt model variants are from the xDeepInt~\cite{yan2020xdeepint}. The key components of the xDeepInt, as illustrated in \autoref{fig_pin_subspace_ftrl}, include Polynomial Interaction Network(PIN) layers, Subspace-crossing Mechanism, and GroupLasso-FTRL/FTRL composite optimization strategy:

\begin{itemize}[leftmargin=10pt]
    \item The PIN layer can capture the higher-order polynomial interactions and is the backbone of our architectures.
    \item The Subspace-crossing Mechanism enables the PIN to learn both vector-wise and bit-wise interactions while controlling the degree of mixture between vector-wise interactions and bit-wise interactions. As shown in the \autoref{fig_subspace}, By the Subspace-crossing Mechanism, One can choose to fully retain the field-level semantic information and only explore vector-wise interaction between features. One can also model the bit-wise interactions by splitting the embedding space into $h$ sub-spaces.
    \item The GroupLasso-FTRL/FTRL composite optimization strategy takes advantage of the properties of different optimizers: GroupLasso-FTRL achieves row-wise sparsity for embedding table, and FTRL achieves element-wise sparsity for weight parameters, as shown in \autoref{fig_ftrl}.
\end{itemize}

The output layer of DynInt model variants is merely a dense layer applied to the feature dimension. The output of PIN is a feature map that consists different degree of feature interactions, including input feature map preserved by residual connections and higher-order feature interactions learned by PIN. The output layer work as:
\begin{equation}
\begin{aligned}
    \hat{y} & = \sigma\big((W_{out}X_{l} + b_{out}\mathbf{1}^T)\mathbf{1} \big)
\end{aligned}
\end{equation}
where $\sigma$ is the sigmoid function, $W_{out}\in R^{1\times F}$ is an aggregation vector that linearly combines all the learned feature interactions, and $b\in R$ is the bias.

We will mainly focus on enhancing the architecture of PIN layer such that the models become instance-aware. Therefore, we will mainly focus on the enhancement of the PIN layer. For simplicity in notations, the subspace-crossing mechanism will not be included in the model introduction and will be included in the implementation for the model performance.

\subsection{Polynomial Interaction Network}
We first go through the architecture of the PIN layer. The mathematical representation of the $l$-th PIN layer with residual connection is given by
\begin{equation}
\begin{aligned}
    X_{l} = X_{l-1}\circ ( W_{l-1} \cdot X_0 ) + X_{l-1}
\end{aligned}
\end{equation}
As \autoref{fig_pin} illustrated, $X_l$ is the $l$-th layer output of PIN. $X_l \in R^{F\times D}$, where $F$ is the number of features, and $D$ is the embedding dimension. $W_{l}\in R^{F\times F}$ is the static parameter that is shared across different instances.

In order to improve the representation power of PIN, we enable the PIN layer parameters to be instance-aware, which enables the dynamic vector-wise and bit-wise interactions modeling. We propose two formulations as follows.

\subsection{Dynamic Activation (DynInt-DA)}
The first scheme is to add dynamic mechanism to the output of each PIN layer.
\begin{equation}
\begin{aligned}
    X_{l} = \Big(X_{l-1}\circ ( W_{l-1} \cdot X_0) \Big) \circ G_{l-1}^{(i)} + X_{l-1}
\end{aligned}
\end{equation}
We multiply the output of the $l$-th layer by a instance-aware gating vector $G_l^{(i)}$. The gating vector $G_l^{(i)}$ can be modeled by a two-layer DNN with reduction ratio $r$, using $X_{0}$ as input.

Empirically, we find that the initialized values of $G_l^{(i)}$ around 1.0 result in better convergence. We set the activation function of gating network as: $\text{Sigmoid}(x) * 2.0$. By using this activation function, we have the $G_l^{(i)}$ initialized around 1.0 when using zero-mean initialization for DNN layer parameters. Meanwhile, we use the input feature map $X_{0}$ as the input of our gating network, which models the dynamic gate for each PIN layer's outputs.

\begin{figure}[htbp]
    \centering
    \includegraphics[width=0.40\textwidth, height=0.30\textwidth]{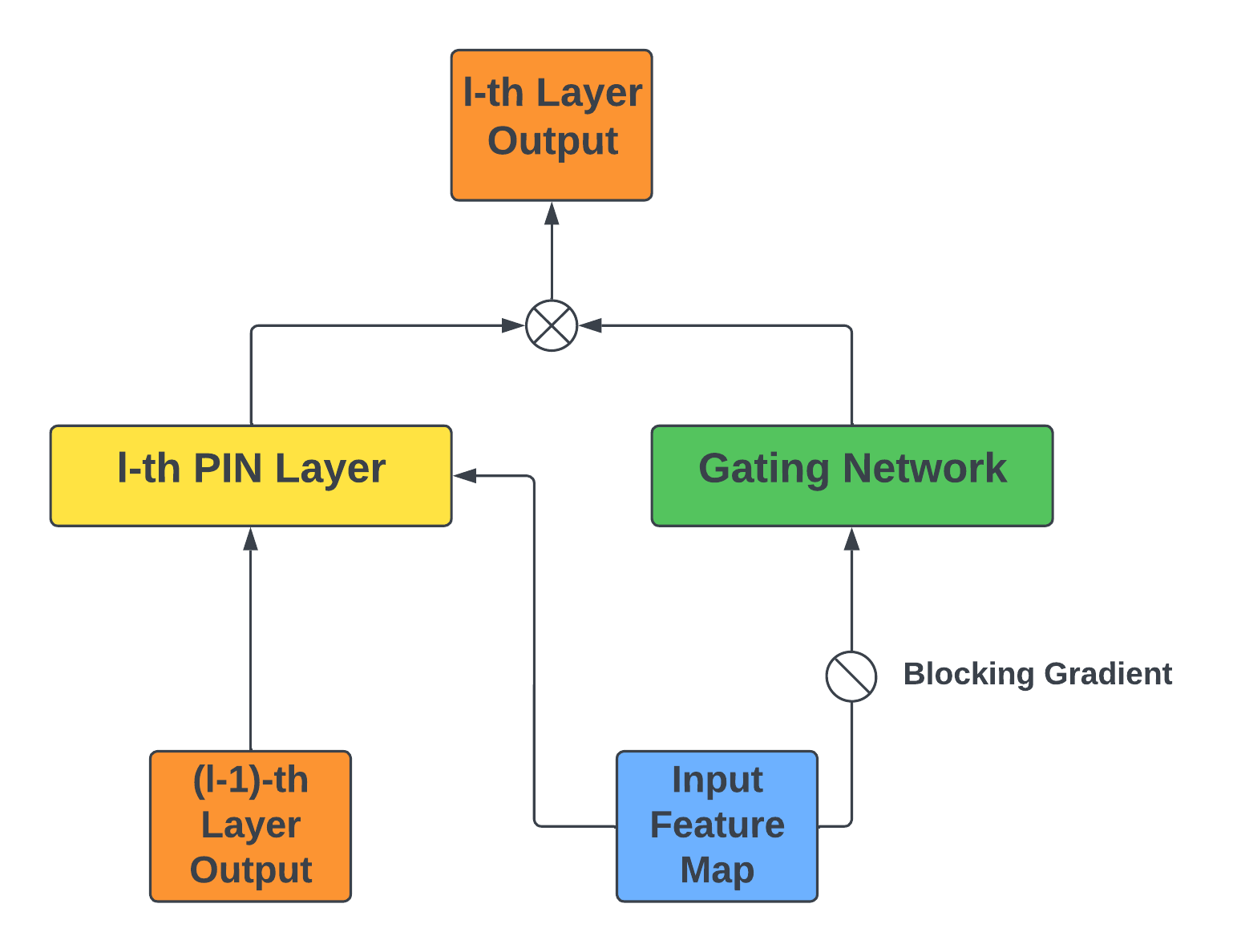}
    \caption{PIN Layer with Dynamic Activation.}
    \label{fig_dynint_da}
	\medskip
    \small
\end{figure}

During the training phase, the input feature map $X_{0}$ does not receive gradients from the gating network to ensure training stability. \autoref{fig_dynint_da} illustrates the architecture of the enhanced PIN layer of the DynInt-DA model.

\subsection{Dynamic Parameters (DynInt-DP)}
Dynamic parameters scheme is to dynamically model $W_{l-1}$ at each PIN layer. We propose two formulations to model the dynamic parameters.

\subsubsection{Dynamic Generated Parameters (DynInt-DGP)}
Instead of using the static parameters, we allow parameter matrices to be data-dependent.
\begin{equation}\label{eq_dgp}
\begin{aligned}
    X_{l} = X_{l-1}\circ ( W^{(i)}_{l-1} \cdot X_0 )+ X_{l-1}
\end{aligned}
\end{equation}
$W^{(i)}_{l-1}$ is a dynamic parameter matrix for the $i$-th instance.

\subsubsection{Dynamic Weighted Parameters (DynInt-DWP)}
Let $G^{(i)}_{l-1}$ be a dynamic weighting matrix for the $i$-th instance. We replace the static parameter matrix $W_{l-1}$ with the Hadamard product of the dynamic weighting matrix $G^{(i)}_{l-1}$ and the static parameter matrix $W_{l-1}$ and obtain that
\begin{equation}\label{eq_dwp}
\begin{aligned}
    X_{l} = X_{l-1}\circ \big( (G^{(i)}_{l-1} \circ  W_{l-1} \cdot X_0 ) \big)+ X_{l-1}
\end{aligned}
\end{equation}

In DGP and DWP, the dynamic parameter matrices $W_{l-1}^{(i)}$ and dynamic weighting matrices $G_{l-1}^{b}$ can also be estimated by two-layer DNNs with reduction ratio $r$. If we take the batch size into consideration, the backward propagation needs to compute the gradients of the matrices of the size $B\times F\times F$, which results in extremely high memory cost. Direct implementations of DWP and DGP are implausible in practice. In the next section, we propose two computation paradigms to implement DGP and DWP with lower memory cost.

\subsection{Low-Rank Approximation for Dynamic Parameters}
The Low-Rank Approximation is commonly used to reduce the memory and computation cost. In PIN layer, we explore the singular values of the trained parameter matrix $W_l$. As shown in \autoref{fig_dynint_dgp_dwp}, the singular values drop dramatically. 

This phenomenon indicates that most of the information is condensed in the top singular values and vectors. Therefore, the low-rank approximation of the parameters can effectively reconstruct the parameters while also improving the serving efficiency.

\begin{figure}[htbp]
    \centering
    \includegraphics[width=0.40\textwidth, height=0.35\textwidth]{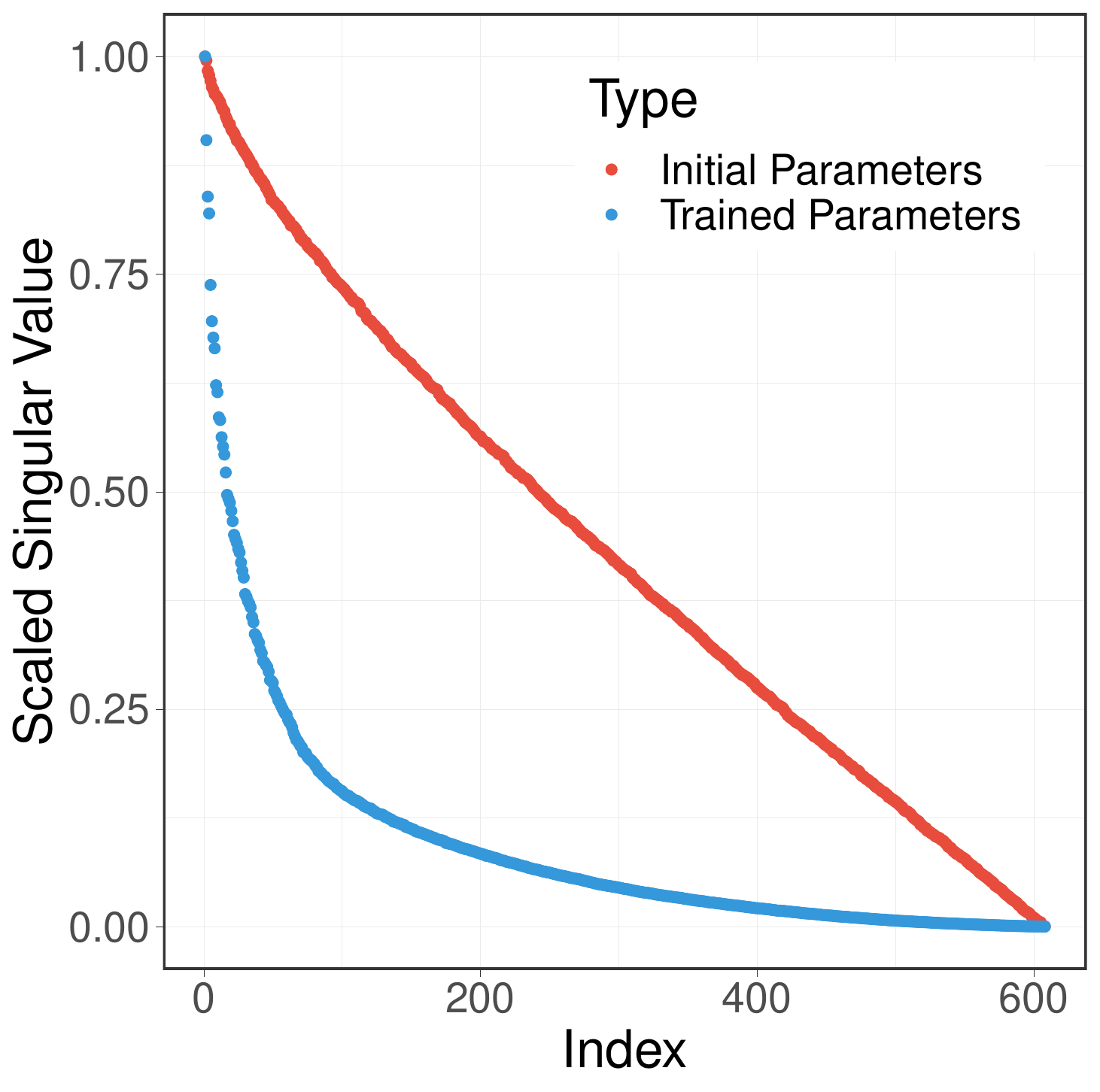}
    \caption{The singular values of the trained parameter matrix $W_l$. We observed that the scaled singular values of the trained static parameter matrices shows a elbow shape, which indicates that the low rank approximation of the parameter matrices preserves the majority of the information.}
    \label{fig_svd_of_wl}
	\medskip
    \small
\end{figure}

\subsection{Computational Paradigms of Dynamic Parameters Formulations}
The computational paradigms assume the dynamic parameter matrix or weighting matrix has low ranks, which controls the model complexity and simplifies the computation. Based on this assumption, we apply singular value decomposition (SVD) to these dynamic matrices.

\subsubsection{Dynamic Generated Parameters}
Suppose we have a rank-K approximation of the dynamic parameter matrix $W_{l-1}^{(i)}$
\begin{equation}
\begin{aligned}
    W_{l-1}^{(i)} = U_{l-1}\cdot\Sigma_{l-1}^{(i)}\cdot V_{l-1}^{T}
\end{aligned}
\end{equation}

where $U_{l-1} \in R^{F\times K}$, $\Sigma_{l-1}^{(i)} \in R^{K\times K}$ and $V_{l-1}\in R^{F\times K}$. We let the singular value matrices be adaptive to instance. Thus all instances will share the same vector space spanned by $U_{l-1}$ and $V_{l-1}$. It can be viewed as a regularization trick. Then $W_{l-1}^{(i)}\cdot X_0$ in \cref{eq_dgp} can be written as:
\begin{equation}
\begin{aligned}
    U_{l-1} \cdot \Sigma_{l-1}^{(i)} \cdot V_{l-1}^{T} \cdot X_0
\end{aligned}
\end{equation}
Since $U_{l-1}$ and $V_{l-1}$ are shared static parameters across all instances. Thus we only need to store two matrices of size $F\times K$ and $B$ diagonal matrices of size $K\times K$ during the forward propagation. Since $\Sigma_{l-1}^{(i)}$ is data-dependent, we will use a two-layer DNN with reduction ratio $r$, with $X_0$ as input to model the non-zero diagonal entries of $\Sigma_{l-1}^{(i)}$. Similar to DynInt-DA, we blocked the gradient of $X_0$ to prevent co-adaption issue.

\autoref{fig_dynint_dgp} shows how we decompose the $W_{l-1}$ with static parameter $U_{l-1}$ and $V_{l-1}$, and instance-aware dynamic parameter $\Sigma_{l-1}^{(i)}$.

\subsubsection{Dynamic Weighted Parameters}
Suppose ${G}^{(i)}_{l-1}$ is the $(l-1)$-th layer weighting matrix has rank K. Then ${G}^{(i)}_{l-1}$ has a SVD of the form 
\begin{equation}
\begin{aligned}
   \mathbf {G}^{(i)}_{l-1} = \sum_{p=1}^{K} \tilde{\mathbf {u}}^{(i)}_{p,l-1} \tilde{\mathbf {v}}^{(i)}_{p,l-1}
\end{aligned}
\end{equation}
where $\tilde{\mathbf {u}}^{(i)}_{p,l-1}=\sqrt{\sigma_{p,l-1}^{(i)}} \mathbf {u}_{p,l-1}^{(i)}$ and $\tilde{\mathbf {v}}_{p,l-1}^{(i)}=\sqrt{\sigma_{p,l-1}^{(i)}} \mathbf {v}^{(i)}_{p,l-1}$. $\tilde{\mathbf {u}}_{p,l-1}^{(i)}$ and $\mathbf{v}_{p,l-1}^{(i)}$ are unit-norm vectors that satisfy that $\tilde{\mathbf {u}}^{(i)}_{p,l-1} \perp \tilde{\mathbf {v}}^{(i)}_{p,l-1}$ for all p's.  Therefore, the PIN layer with dynamic parameters can be re-written as
\begin{equation}
\begin{aligned}
    X_{l} &= \sum_{p=1}^{K} X_{l-1}\circ \Big( \big( \textbf{diag}(\tilde{\mathbf {u}}^{(i)}_{p,l-1}) \cdot W_{l-1} \cdot \textbf{diag}(\tilde{\mathbf {v}}^{(i)}_{p,l-1}) \big) \cdot X_0 \Big) + X_{l-1}
\end{aligned}
\end{equation}
where $\tilde{\mathbf {u}}^{(i)}_{p,l-1} \perp \tilde{\mathbf {v}}^{(i)}_{p,l-1}$ for all p's. We simplify the Hadamard product to the matrix product. And $W_{l-1}$ is the shared parameter matrix in this formulation. $\tilde{\mathbf {u}}^{(i)}_{p,l-1}$ and $\tilde{\mathbf {v}}^{(i)}_{p,l-1}$  are the personalized weights. Since we can simplify the hadmard product of a rank-1 matrix and a matrix with a row-wise broadcasting operation and a column-wise broadcasting operation, we apply the vectorized form to the DWP. In each forward propagation iteration, $K$ pairs $\tilde{\mathbf {u}}^{(i)}_{p,l-1}$, $\tilde{\mathbf {v}}^{(i)}_{p,l-1}$ and a shared parameter matrix $W_{l-1}$ will be stored.

\begin{figure}[htbp]
\centering
    \begin{subfigure}[b]{0.50\textwidth}  
    \includegraphics[width=\textwidth]{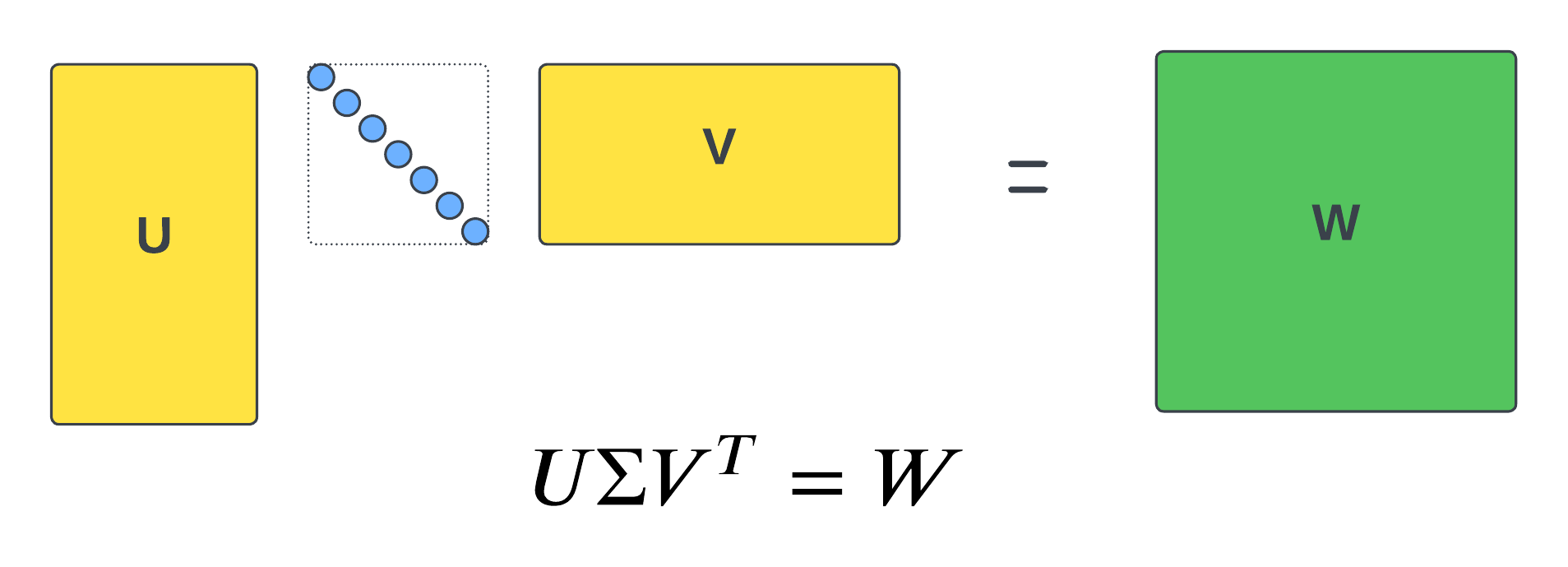}
    \caption{Low rank approximations of dynamic parameter matrices $W$. In implementation, we let $U$ and $V$ to be static parameters. $\Sigma$ is dynamic parameters.}
    \label{fig_dynint_dgp}
    \end{subfigure}
    \vfill
    \begin{subfigure}[b]{0.50\textwidth}  
    \includegraphics[width=\textwidth]{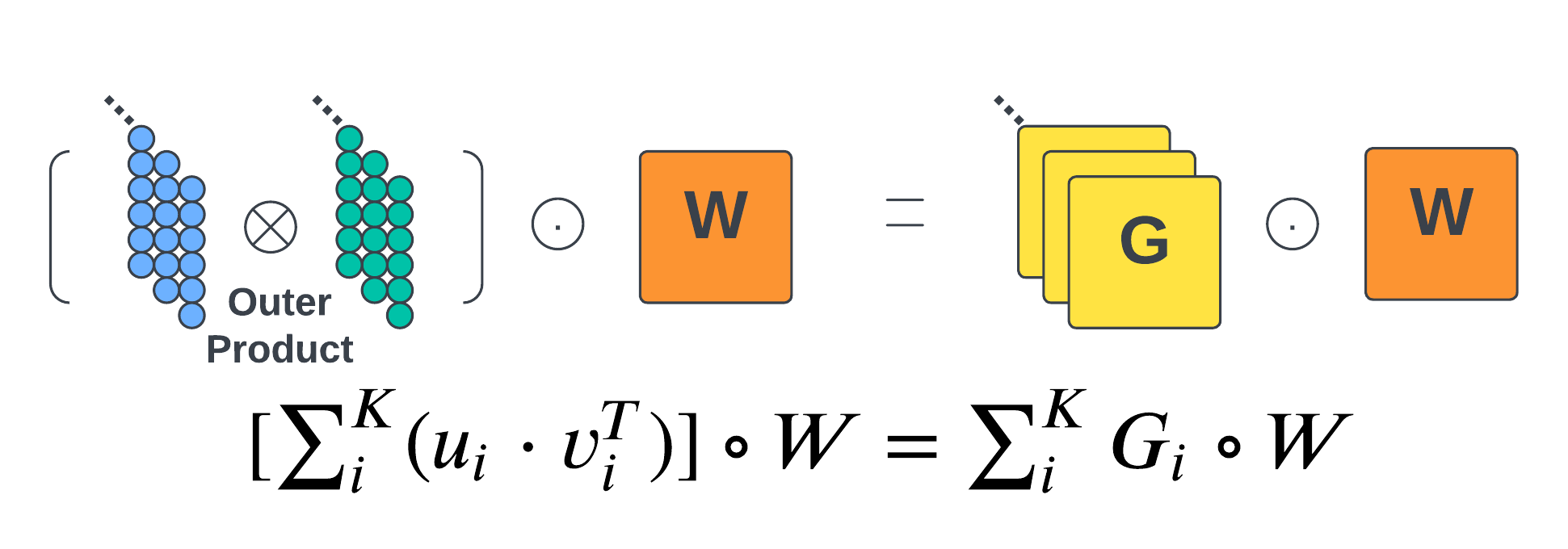}
    \caption{Low rank approximations of the dynamic weighting matrices $G$. In implementation, we let $u_i$ and $v_i$ to be dynamic parameters.}
    \label{fig_dynint_dwp}
    \end{subfigure}
\caption{Low rank approximation for two dynamic parameters formulation. Both $DGP$ and $DWP$ use SVD to approximate the dynamic parameter matrices(DGP) or dynamic weighting matrices(DWP).}
\label{fig_dynint_dgp_dwp}
\end{figure}

\autoref{fig_dynint_dwp} illustrates the decomposition of dynamic weighting matrix $\mathbf {G}^{(i)}_{l-1}$. The $K$ pairs $\tilde{\mathbf {u}}^{(i)}_{p,l-1}$, $\tilde{\mathbf {v}}^{(i)}_{p,l-1}$ are modeled by a two-layer DNN network with reduction ratio $r$ and using gradient blocked $X_{0}$ as input, similar to aforementioned setup. Since we want to ensure the entries of $\mathbf {G}^{(i)}_{l-1}$ is around 1.0 after initilization, we add $1.0 / K$ to the output of the DNN that modeling $K$ pairs $\tilde{\mathbf {u}}^{(i)}_{p,l-1}$, $\tilde{\mathbf {v}}^{(i)}_{p,l-1}$. We find this trick improves the training stability.

\subsubsection{Memory Cost Reduction}
For both DynInt-DWP and DynInt-DGP, the memory cost for the instance-aware parameter matrix is $B\times F\times F$. For the DynInt-DGP with the proposed computation paradigm, the memory cost of the low-rank approximation of the parameter matrix is $B \times K \times K$ + $2\times F\times K$ with $K << F$. If $K = \frac{1}{10}F$, the memory cost for storing the personalized weighted matrix is reduced by $99\%$ with the proposed computation paradigm. For the the DynInt-DWP with the proposed computation paradigm, the memory cost of the low-rank approximation of the parameter matrix is $2\times B \times F \times K + F \times F$ with $K << F$. If $K = \frac{1}{10}F$, the memory cost for storing the personalized weighted matrix is reduced by $80\%$. 

\subsubsection{FLOPs Analysis}
DynInt-DGP requires $O(F\times F\times D)$ FLOPs for each instance. With SVD, the computational cost can be reduced to $O(2\times F\times K\times K + F\times K\times D)$, the flops of SVD is significantly smaller than $O(F\times F\times D)$ given $K << D, F$. DynInt-DWP also requires $O(F\times F\times D)$ FLOPs for each instance. With SVD, the computational cost is $O(2K\times F\times F + K\times F\times F\times D)$ which is comparable with $O(F\times F\times D)$ when $K << D, F$ 

Therefore, the proposed computation paradigms can significantly reduce the memory cost. DGP can also reduce the computational cost with the proposed computational paradigm.

\subsection{Orthogonal Regularization and Loss Function}
Both computational paradigms are based on low-rank approximations and SVD. While we model the instance-aware dynamic parameters, each instance also shares certain model parameters. This helps to control the model complexity.

The orthogonality is important to guarantee performance. We think there are two reasons. First, it is required in by SVD. We need the singular vectors to be orthogonal. Second, the SVD decomposed DGP and DWP can be viewed as ensembles of multiple rank-1 DGP and DWP.

The orthogonality constraint guarantees that the column space spanned by each singular vector are distinct from each other. Therefore, we add the orthogonal penalties to the loss functions to encourage the model to learn diverse fine-grained representations and achieve better performance.

\subsubsection{DynInt-DGP Loss}
\begin{equation}\label{loss_dgp}
\begin{aligned}
     Loss & = \sum_{i=1}^N \big(y_i log(\hat{y}_i) +  (1-y_i)log(1-\hat{y}_i) \big) \\
   & + \lambda \sum_{i=1}^N \sum_{l=1}^L\sum_{p>q}^K \|\frac{\tilde{\mathbf {u}}_{p,l-1}^T\tilde{\mathbf {u}}_{q,l-1}}{\| \tilde{\mathbf {u}}_{p,l-1} \| \cdot \| \tilde{\mathbf {u}}_{q,l-1} \|}\|_1 + \|\frac{\tilde{\mathbf {v}}_{p,l-1}^T\tilde{\mathbf {v}}_{q,l-1}}{\|\tilde{\mathbf {v}}_{p,l-1}\|\cdot\|\tilde{\mathbf {v}}_{q,l-1}\|}\|_1
\end{aligned}
\end{equation}
For DGP, we add the sum of the $l_1$ norm of the of cosine similarity of $\tilde{\mathbf{u}}_{\cdot,l}$ pairs and $\tilde{\mathbf{v}}_{\cdot,l}$ pairs across different PIN layers as penalty. The penalty term in \cref{loss_dgp} ensures the orthogonality between $\tilde{\mathbf {u}}_{\cdot,l}$ pairs and $\tilde{\mathbf {v}}_{\cdot,l}$ pairs.
\subsubsection{DynInt-DWP Loss}
\begin{equation}\label{loss_dwp}
\begin{aligned}
   Loss & = \sum_{i=1}^N \big(y_i log(\hat{y}_i) +  (1-y_i)log(1-\hat{y}_i)
   \big)\\ 
   & + \lambda \sum_{i=1}^N\sum_{l=1}^L\sum_{p>q}^K \|\frac{\tilde{\mathbf {u}}^{(i)}_{p,l-1} \tilde{\mathbf {u}}^{(i)}_{q,l-1}}{\| \tilde{\mathbf {u}}^{(i)}_{p,l-1} \| \cdot \| \tilde{\mathbf {u}}^{(i)}_{q,l-1} \|}\|_1 + \|\frac{\tilde{\mathbf {v}}^{(i)}_{p,l-1} \tilde{\mathbf {v}}^{(i)}_{q,l-1}}{\|\tilde{\mathbf {v}}^{(i)}_{p,l-1}\|\cdot\|\tilde{\mathbf {v}}^{(i)}_{q,l-1}\|}\|_1
\end{aligned}
\end{equation}
For DWP, we add the sum of the $l_1$ norm of the cosine similarity of $\tilde{\mathbf{u}}^{(i)}_{\cdot,l}$ pairs and $\tilde{\mathbf{v}}^{(i)}_{\cdot,l}$ pairs across different PIN layers and instances. The penalty term in \cref{loss_dwp} ensures the orthogonality between $\tilde{\mathbf {u}}^{(i)}_{\cdot,l}$ pairs and $\tilde{\mathbf {v}}^{(i)}_{\cdot,l}$ pairs.

\section{Experiments}
In this section, we focus on evaluating the effectiveness of our proposed models and seeking answers to the following research questions::
\begin{itemize}[leftmargin=10pt]
	\item \textbf{Q1}: How do our proposed DynInt model variants perform compared to each baseline in CTR prediction problem?
	\item \textbf{Q2}: How do our proposed DynInt model variants perform compared to each baseline, which only learns static feature interactions?
	\item \textbf{Q3}: How do different hyper-parameter settings influence the performance of DynInt variants?
	\item \textbf{Q4}: How does the orthogonal regularization help the effectiveness of DynInt?
\end{itemize}

\subsection{Experiment Setup}

\subsubsection{Datasets}
We evaluate our proposed model on three public real-world datasets widely used for research.

\textbf{1. Criteo.}\footnote{https://www.kaggle.com/c/criteo-display-ad-challenge} Criteo dataset is from Kaggle competition in 2014. Criteo AI Lab officially released this dataset after, for academic use. This dataset contains 13 numerical features and 26 categorical features. We discretize all the numerical features to integers by transformation function $\lfloor Log\left(V^{2}\right) \rfloor$ and treat them as categorical features, which is conducted by the winning team of Criteo competition.

\textbf{2. Avazu.}\footnote{https://www.kaggle.com/c/avazu-ctr-prediction} Avazu dataset is from Kaggle competition in 2015. Avazu provided 10 days of click-through data. We use 21 features in total for modeling. All the features in this dataset are categorical features.

\textbf{3. iPinYou.}\footnote{http://contest.ipinyou.com/} iPinYou dataset is from iPinYou Global RTB(Real-Time Bidding) Bidding Algorithm Competition in 2013. We follow the data processing steps of~\cite{zhang2014real} and consider all 16 categorical features.

For all the datasets, we randomly split the examples into three parts: 70\% is for training, 10\% is for validation, and 20\% is for testing. We also remove each categorical features' infrequent levels appearing less than 20 times to reduce sparsity issue. Note that we want to compare the effectiveness and efficiency on learning higher-order feature interactions automatically, so we do not do any feature engineering but only feature transformation, e.g., numerical feature bucketing and categorical feature frequency thresholding.

\subsubsection{Evaluation Metrics}
We use AUC and LogLoss to evaluate the performance of the models.

\textbf{LogLoss} LogLoss is both our loss function and evaluation metric. It measures the average distance between predicted probability and true label of all the examples.

\textbf{AUC} Area Under the ROC Curve (AUC) measures the probability that a randomly chosen positive example ranked higher by the model than a randomly chosen negative example. AUC only considers the relative order between positive and negative examples. A higher AUC indicates better ranking performance.

\subsubsection{Competing Models}
We compare all of our DynInt variants with following models: LR (Logistic Regression)~\cite{mcmahan2011follow,mcmahan2013ad}, FM (Factorization Machine)~\cite{rendle2010factorization}, DNN (Multilayer Perceptron), Wide \& Deep~\cite{cheng2016wide}, DeepCrossing~\cite{shan2016deep}, DCN (Deep \& Cross Network)~\cite{wang2017deep}, DCN V2~\cite{wang2021dcn}, PNN (with both inner product layer and outer product layer)~\cite{qu2016product,qu2018product}, DeepFM~\cite{guo2017deepfm}, xDeepFM~\cite{lian2018xdeepfm}, AutoInt~\cite{song2018autoint} and FiBiNET~\cite{huang2019fibinet}. Some of the models are state-of-the-art models for CTR prediction problem and are widely used in the industry.

\subsubsection{Reproducibility}
We implement all the models using Tensorflow~\cite{abadi2016tensorflow}. The mini-batch size is 4096, and the embedding dimension is 16 for all the features. For optimization, we employ Adam~\cite{kingma2014adam} with learning rate is tuned from $10^{-4}$ to $10^{-3}$ for all the neural network models, and we apply FTRL~\cite{mcmahan2011follow,mcmahan2013ad} with learning rate tuned from $10^{-2}$ to $10^{-1}$ for both LR and FM. For regularization, we choose L2 regularization with $\lambda$ ranging from $10^{-4}$ to $10^{-3}$ for dense layer. Grid-search for each competing model's hyper-parameters is conducted on the validation dataset. The number of dense or interaction layers is from 1 to 4. The number of neurons ranges from 128 to 1024. All the models are trained with early stopping and are evaluated every 2000 training steps.

Similar to \cite{yan2020xdeepint}, the setup is as follows for the hyper-parameters search of xDeepInt and DynInt: The number of recursive feature interaction layers is searched from 1 to 4. For the number of sub-spaces $h$, the searched values are 1, 2, 4, 8 and 16. Since our embedding size is 16, this range covers from complete vector-wise interaction to complete bit-wise interaction. For the reduction ratio of all the DynInt variants, we search from 2 to 16. The kernel size for DynInt-DGP is tuned from 16 to 128. The latent rank of weighting matrix for DynInt-DWP is searched from 1 to 4. We use G-FTRL optimizer for embedding table and FTRL for Dynamic PIN layers with learning rate tuned from $10^{-2}$ to $10^{-1}$.

\subsection{Model Performance Comparison (Q1)}

\begin{table}[H]
	\caption{Performance Comparison of Different Algorithms on Criteo, Avazu and iPinYou Dataset.}\label{tbl_model_performance}
	\centering
	\resizebox{1.0\linewidth}{!}{
		\begin{tabular}{ccccccc}
			\hline
			& \multicolumn{2}{c}{Criteo}        & \multicolumn{2}{c}{Avazu}         & \multicolumn{2}{c}{iPinYou}                      \\
			Model        & AUC             & LogLoss         & AUC             & LogLoss         & AUC             & LogLoss           \\
			\hline
			LR           & 0.7924          & 0.4577          & 0.7533          & 0.3952          & 0.7692          & 0.005605          \\
			FM           & 0.8030          & 0.4487          & 0.7652          & 0.3889          & 0.7737          & 0.005576          \\
			DNN          & 0.8051          & 0.4461          & 0.7627          & 0.3895          & 0.7732          & 0.005749          \\
			Wide\&Deep   & 0.8062          & 0.4451          & 0.7637          & 0.3889          & 0.7763          & 0.005589          \\
			DeepFM       & 0.8069          & 0.4445          & 0.7665          & 0.3879          & 0.7749          & 0.005609          \\
			DeepCrossing & 0.8068          & 0.4456          & 0.7628          & 0.3891          & 0.7706          & 0.005657          \\
			DCN          & 0.8056          & 0.4457          & 0.7661          & 0.3880          & 0.7758          & 0.005682          \\
			PNN          & 0.8083          & 0.4433          & 0.7663          & 0.3882          & 0.7783          & 0.005584          \\
			xDeepFM      & 0.8077          & 0.4439          & 0.7668          & 0.3878          & 0.7772          & 0.005664          \\
			AutoInt      & 0.8053          & 0.4462          & 0.7650          & 0.3883          & 0.7732          & 0.005758          \\
			FiBiNET      & 0.8082          & 0.4439          & 0.7652          & 0.3886          & 0.7756          & 0.005679          \\
			DCN V2       & 0.8086          & 0.4433          & 0.7662          & 0.3882          & 0.7765          & 0.005593          \\
			xDeepInt     & 0.8111          & 0.4408          & 0.7672          & 0.3876          & 0.7790          & 0.005567          \\
			\hline
			DynInt-DA    & 0.8125          & 0.4398          & 0.7677          & 0.3871          & 0.7802          & 0.005552          \\
			DynInt-DGP   & 0.8127          & 0.4397          & \textbf{0.7686} & \textbf{0.3867} & \textbf{0.7813} & \textbf{0.005547} \\
			DynInt-DWP   & \textbf{0.8132} & \textbf{0.4393} & 0.7671          & 0.3878          & 0.7798          & 0.005564          \\
			\hline
	\end{tabular}}
\end{table}

The overall performance of different model architectures is listed in \Cref{tbl_model_performance}. We have the following observations in terms of model effectiveness:
\begin{itemize}[leftmargin=10pt]
	\item FM brings the most significant relative boost in performance while we increase model complexity, compared to LR baseline. This reveals the importance of learning explicit vector-wise feature interactions.
	\item Models that learn vector-wise and bit-wise interactions simultaneously consistently outperform other models. This phenomenon indicates that both types of feature interactions are essential to prediction performance and compensate for each other.
	\item xDeepInt achieves the best prediction performance among all static interaction learning models. The superior performance could attribute to the fact that xDeepInt models the bounded degree of polynomial feature interactions by adjusting the depth of PIN and achieve different complexity of bit-wise feature interactions by changing the number of sub-spaces.
	\item xDeepInt and all the DynInt model variants achieve better performance compared to other baseline models without integrating with DNN for learning implicit interactions from feature embedding, which indicates that the well-approximated polynomial interactions can potentially overtake DNN learned implicit interactions under extreme high-dimensional and sparse data settings.
	\item All the DynInt model variants achieve better performance compared to all static interaction learning models, which indicates that learning dynamic interactions can further improve the model performance. While different datasets favor a different type of dynamic interaction modeling approach, that's mainly driven by the fact that DynInt model variants have different model capacity and regularization effects by design. We defer a detailed discussion of these models in later sections.
\end{itemize}

\subsection{Feature Interaction Layer Comparison (Q2)}
We evaluate each baseline model \textbf{without integrating with DNN} on Criteo dataset, which means only the feature interaction layer will be applied. The overall performance of different feature interaction layers is listed in \Cref{tbl_interaction_layer_performance}. We can observe that most of the baseline model initially integrated with DNN, except DCN V2, has a significant performance drop. xDeepInt and DynInt variants were not integrating with DNN for learning implicit interactions and maintained the original performance. Based on the above observations, we developed the following understandings:

\begin{itemize}[leftmargin=10pt]
	\item Most of the baseline models rely on DNN heavily to compensate for the feature interaction layers, and the overall performance is potentially dominated by the DNN component.
	\item For model architectures like DeepFM and PNN, their interaction layer can only learn two-way interactions, so they have inferior performance compared to DCN v2, xDeepInt and DynInt variants.
	\item DCN v2, xDeepInt, and DynInt variants only learn polynomial interactions while still achieving better performance than other baseline models, which indicates that higher-order interactions do exist in dataset, and polynomial interactions are essential to the superior performance.
\end{itemize}

\begin{table}[H]
	\caption{Performance Comparison of Different Feature Interaction Layer on Criteo Dataset.}\label{tbl_interaction_layer_performance}
	\centering
	\begin{tabular}{c|cc}
		\hline
		               & AUC     & LogLoss \\
		\hline
		DeepFM         & 0.8030  & 0.4487  \\
		DCN            & 0.7941  & 0.4566  \\
		PNN            & 0.8066  & 0.4451  \\
		xDeepFM        & 0.8065  & 0.4454  \\
		AutoInt        & 0.8056  & 0.4458  \\
		DCN V2         & 0.8082  & 0.4436  \\
		xDeepInt       & 0.8111  & 0.4408  \\
		DynInt-DA      & 0.8125  & 0.4398  \\
		DynInt-DGP     & 0.8127  & 0.4397  \\
		DynInt-DWP     & 0.8132  & 0.4393  \\
		\hline
	\end{tabular}
\end{table}

\subsection{Hyper-Parameter Study (Q3 and Q4)}
In order to have deeper insights into the proposed model, we conduct experiments on three datasets and compare several variants of DynInt on different hyper-parameter settings. This section evaluates the model performance change with respect to hyper-parameters that include: 1) depth of dynamic PIN layers; 2) number of sub-spaces; 3) kernel size of generated parameters for DynInt-DGP; 4) latent rank of weighting matrix for DynInt-DWP; 5) reduction ratio; 6) orthogonal regularization rate for DynInt-DGP and DynInt-DWP

\subsubsection{Depth of Network}
The depth of dynamic PIN layers determines the order of feature interactions learned. \Cref{tbl_hyper_parameter_depth} illustrates the performance change with respect to the number of layers. When the number of layers is set to 0, our model is equivalent to logistic regression and no interactions are learned. In this experiment, we set the number of sub-spaces as 1, to disable the bit-wise feature interactions. Since we are mainly checking the performance v.s. the depth change, we average the performance of DynInt-DA, DynInt-DGP and DynInt-DWP for simplicity.

\begin{table}[H]
    \caption{Impact of hyper-parameters: Number of Layers}\label{tbl_hyper_parameter_depth}
    \footnotesize
    \centering
    \resizebox{\linewidth}{!}
    {\begin{tabular}{ccc|cccccc}
        \hline
        &Dataset       &\#Layers       & 0                 & 1                 & 2                 & 3                 & 4                 & 5        \\
        \hline
        &Criteo        &AUC            & 0.7921           & 0.8058             & 0.8068            & 0.8075            & \textbf{0.8085}   & 0.8081   \\
        &              &LogLoss        & 0.4580           & 0.4455             & 0.4445            & 0.4440            & \textbf{0.4432}   & 0.4436   \\
        \hline
        &Avazu         &AUC            & 0.7547           & 0.7664             & 0.7675            & \textbf{0.7682}   & 0.7678            & 0.7670   \\
        &              &LogLoss        & 0.3985           & 0.3882             & 0.3870            & \textbf{0.3868}   & 0.3872            & 0.3877   \\
        \hline
        &iPinYou       &AUC            & 0.7690           & 0.7749             & 0.7787            & \textbf{0.7797}   & 0.7785            & 0.7774   \\
        &              &LogLoss        & 0.005604         & 0.005575           & 0.005569          & \textbf{0.005565} & 0.005578          & 0.005582 \\
        \hline
    \end{tabular}}
\end{table}

\subsubsection{Number of Sub-spaces}
The subspace-crossing mechanism enables the proposed model to control the complexity of bit-wise interactions. \Cref{tbl_hyper_parameter_subspace} demonstrates that subspace-crossing mechanism boosts the performance. In this experiment, we set the number of PIN layers as 3, which is generally a good choice but not necessarily the best setting for each dataset. Since we are mainly checking the performance v.s. the number of sub-spaces change, we average the performance of DynInt-DA, DynInt-DGP and DynInt-DWP for simplicity.

\begin{table}[H]
    \caption{Impact of hyper-parameters: Number of Sub-Spaces}\label{tbl_hyper_parameter_subspace}
    \footnotesize
    \centering
    \resizebox{\linewidth}{!}
    {\begin{tabular}{ccc|cccccc}
        \hline
        &Dataset       &\#Sub-spaces & 1                 & 2                 & 4                 & 8                 & 16              \\
        \hline
        &Criteo        &AUC          & 0.8079            & 0.8085            & 0.8092            & 0.8107            & \textbf{0.8125} \\
        &              &LogLoss      & 0.4439            & 0.4433            & 0.4421            & 0.4408            & \textbf{0.4397} \\
        \hline
        &Avazu         &AUC          & 0.7670            & 0.7677            & \textbf{0.7684}   & 0.7678            & 0.7672          \\
        &              &LogLoss      & 0.3875            & 0.3872            & \textbf{0.3867}   & 0.3872            & 0.3873          \\
        \hline
        &iPinYou       &AUC          & 0.7782            & 0.7792            & \textbf{0.7796}   & 0.7790            & 0.7784          \\
        &              &LogLoss      & 0.005579          & 0.005568          & \textbf{0.005565} & 0.005568          & 0.005582        \\
        \hline
    \end{tabular}}
\end{table}

\subsubsection{Kernel Size of Dynamic Generated Parameter for DynInt-DGP}
The kernel size of DynInt-DGP controls both the number of static parameters and the number of dynamic parameters for DynInt-DGP.

\autoref{fig_kernel_size} shows the performance v.s. the kernel size of the parameter matrix of DynInt-DGP on Criteo dataset. We observe that the performance keeps increasing until we increase the kernel size up to 64. This aligns with our understanding of the performance v.s. model complexity, which also means that the kernel size can be served as a regularization hyper-parameter for approximating the underlying true weight.

\subsubsection{Latent Rank of Dynamic Weighting Parameter for DynInt-DWP}
The latent rank of DynInt-DWP controls the complexity of the dynamic weighting matrix for the shared static parameters.

\begin{figure}[htbp]
\centering
    \begin{subfigure}[b]{0.23\textwidth}  
    \includegraphics[width=\textwidth]{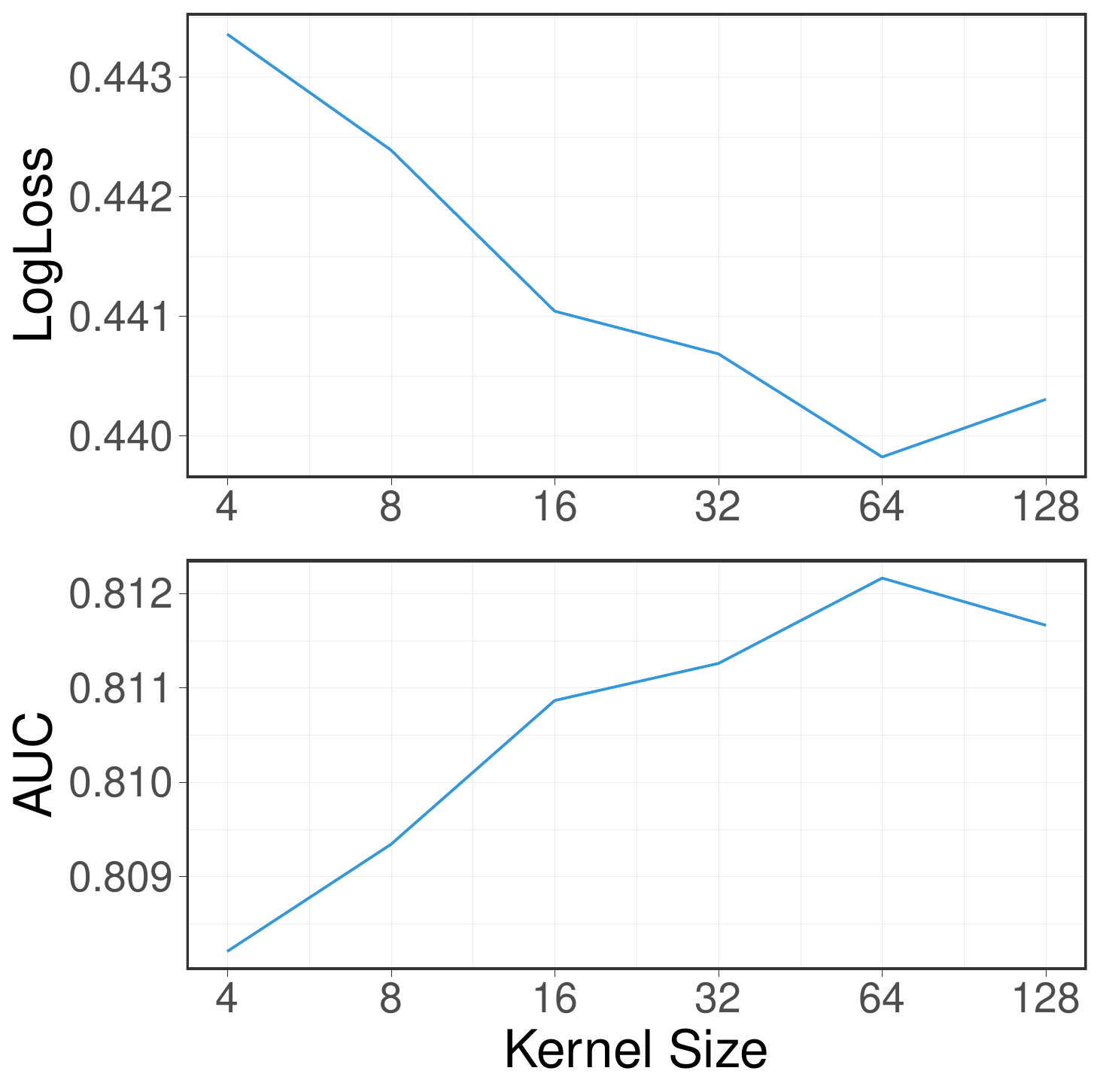}
    \caption{Kernel Size of DynInt-DGP}
    \label{fig_kernel_size}
    \end{subfigure}
    \hfill
    \begin{subfigure}[b]{0.23\textwidth}  
    \includegraphics[width=\textwidth]{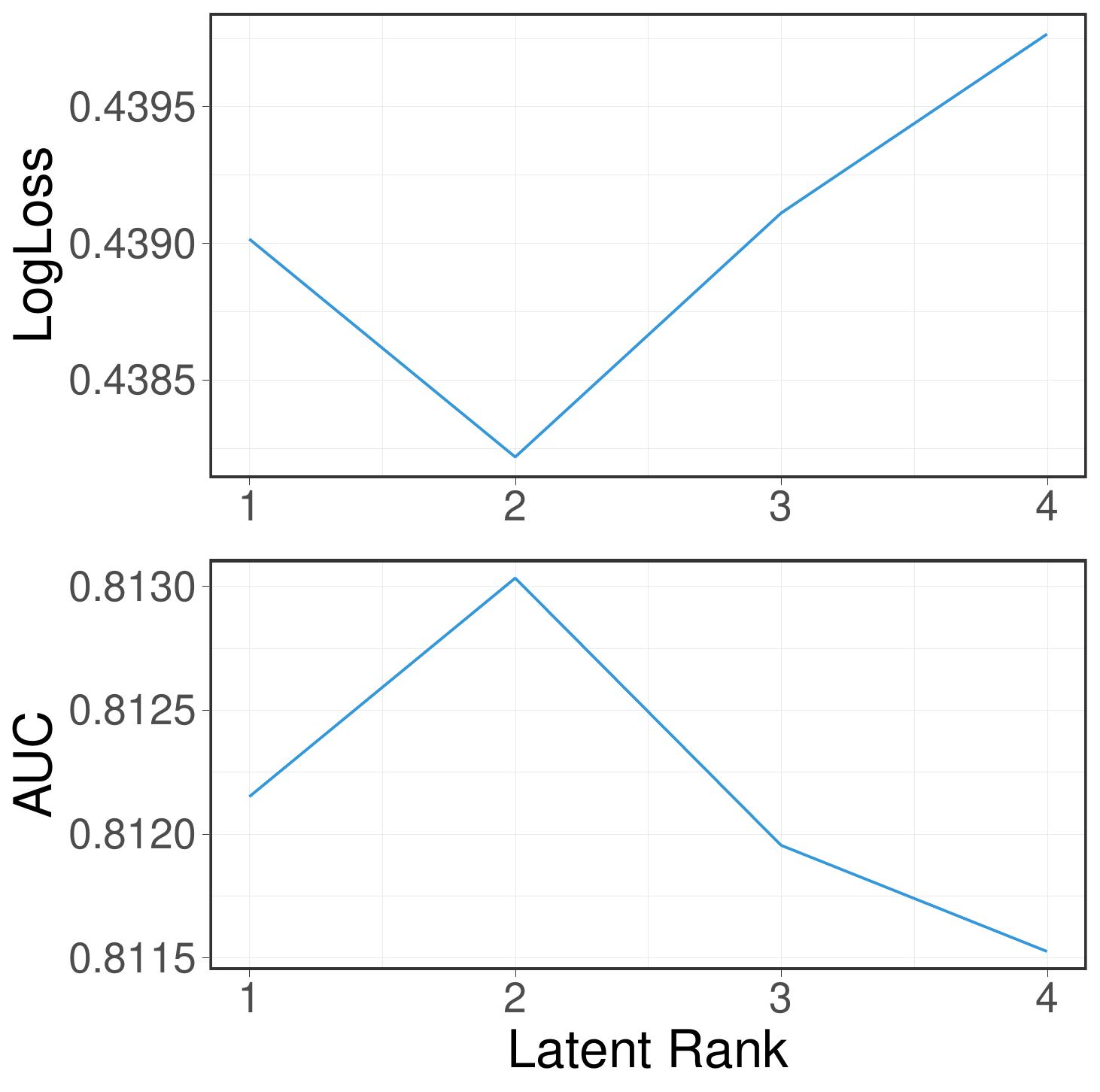}
    \caption{Rank of DynInt-DWP}
    \label{fig_latent_rank}
    \end{subfigure}
\caption{Logloss and AUC v.s. kernel size and latent rank.}
\label{fig_kernel_size_rank}
\end{figure}

\autoref{fig_latent_rank} shows the performance v.s. the latent rank of parameter matrix of DynInt-DGP on Criteo dataset. The complexity of the DynInt-DWP grows fast when we increase the latent rank of the weighting matrix.

\subsubsection{Reduction Ratio}
The reduction ratio is a hyper-parameter that allows us to control the capacity of our network for dynamic parameter weighting/generation. We examine the impact of reduction ratio on Criteo dataset. The correlation between the reduction ratio and performance of different architecture is similar. Other hyper-parameters maintain the same as the best setting of each specific architecture. Here we average their performance for a more straightforward illustration, as each method show a similar trend in our experiment.

\autoref{fig_reduction_ratio} shows the relationship between the different setup of reduction ratio and performance. We observe that the performance is robust to a reasonable range of reduction ratios. Increased complexity does not necessarily improve the performance, while a larger reduction ratio improves the model's efficiency.

\subsubsection{Orthogonal Regularization Rate}
The orthogonal regularization rate is a hyper-parameter for controlling the representation diversity among column spaces of the dynamic parameters. Since the dynamic generated/weighting matrices can be seen as the sum of low-rank matrices, so each decomposed subspace should learn relatively distinct features. We examine the impact of orthogonal regularization rate on Criteo dataset. We average both methods' performance for simplicity, as they show a similar trend in our experiment.

\begin{figure}[htbp]
\centering
    \begin{subfigure}[b]{0.23\textwidth}
    \includegraphics[width=\textwidth]{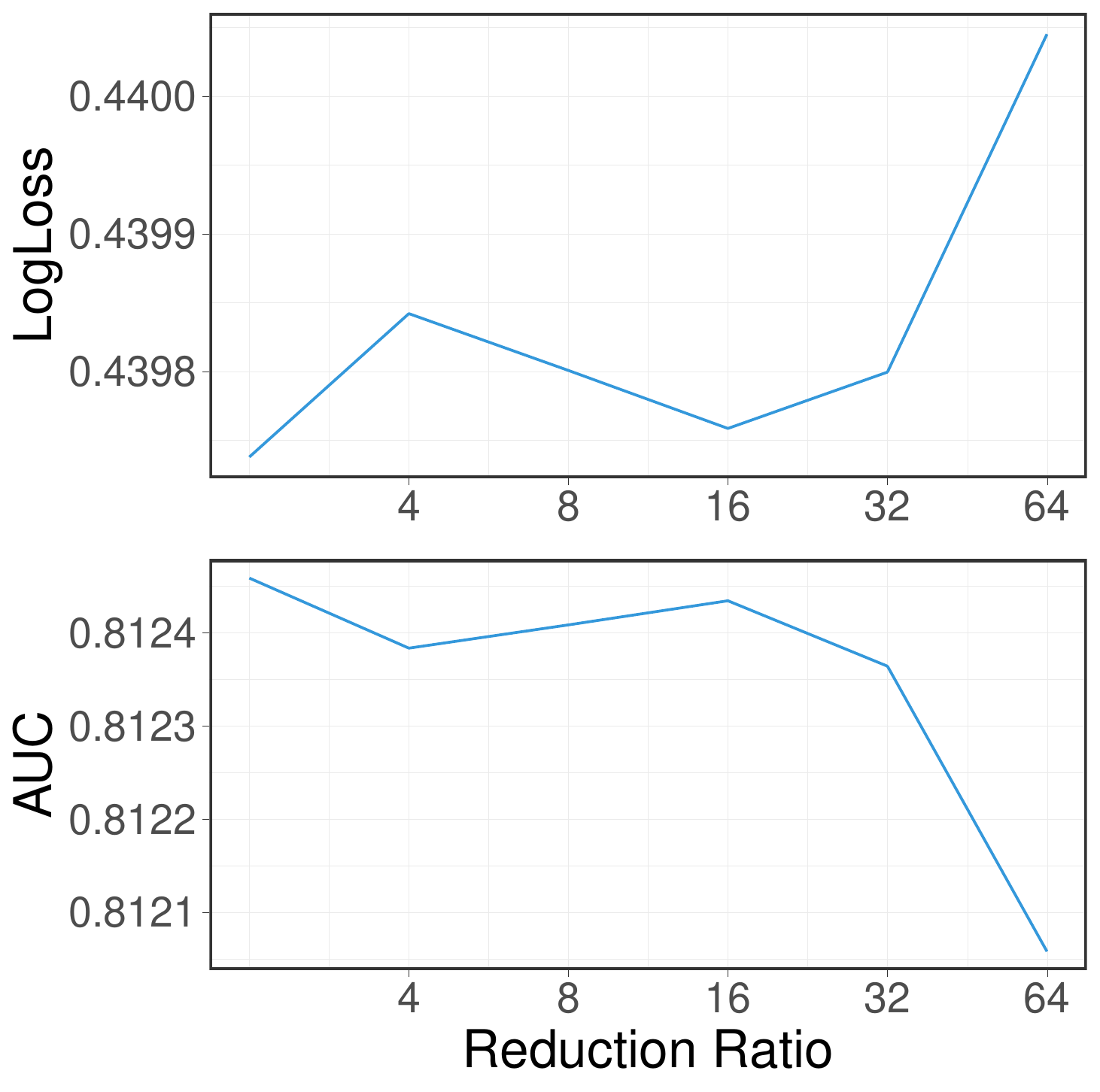}
    \caption{Reduction Ratio}
    \label{fig_reduction_ratio}
    \end{subfigure}
    \hfill
    \begin{subfigure}[b]{0.23\textwidth}  
    \includegraphics[width=\textwidth]{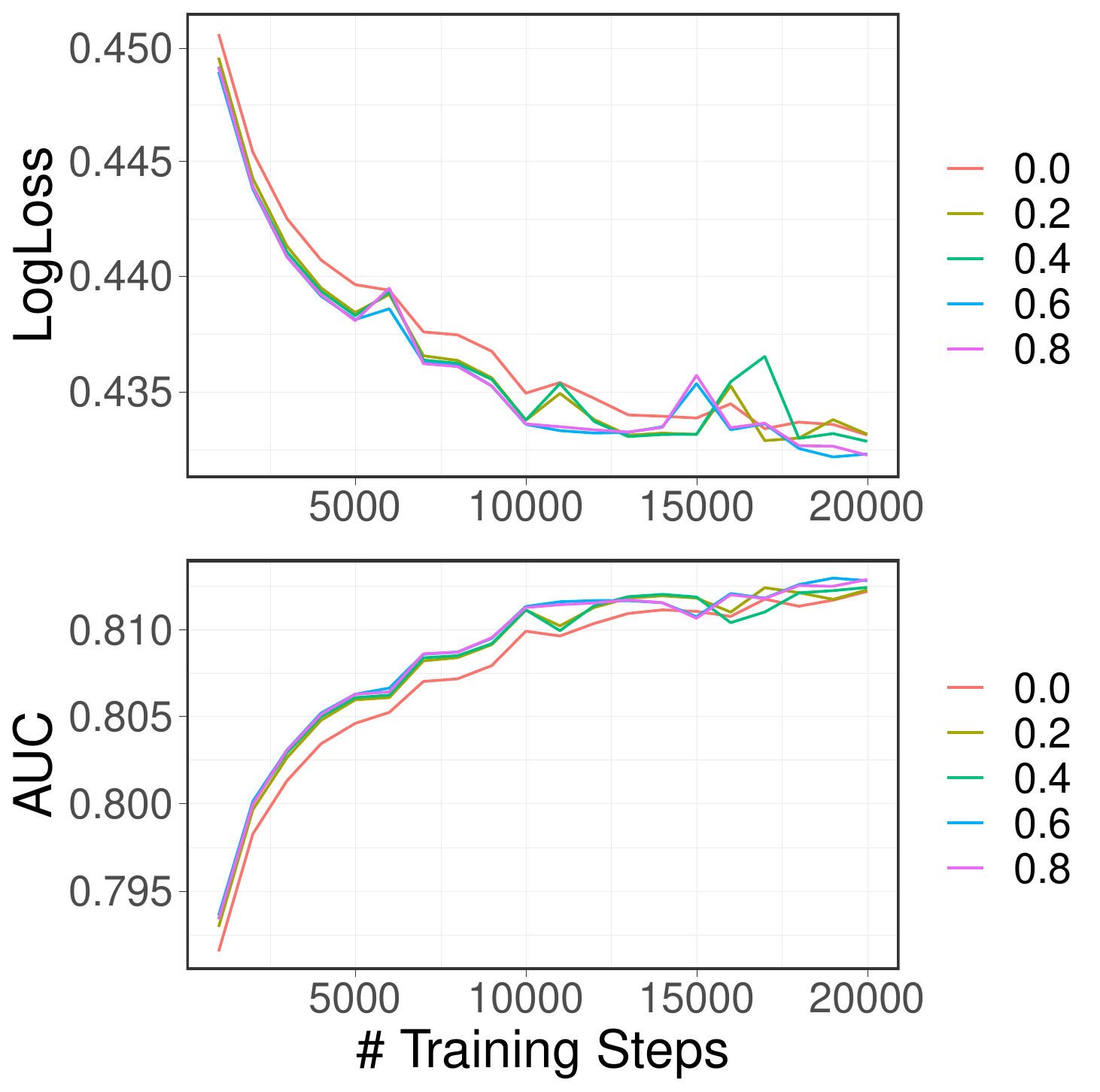}
    \caption{Orthogonal Regularization}
    \label{fig_orthogonal_regularization}
    \end{subfigure}
\caption{Logloss and AUC v.s. reduction ratio and orthogonal regularization rate.}
\label{fig_reduction_ratio_orthogonal_regularization}
\end{figure}

\autoref{fig_orthogonal_regularization} illustrates the performance v.s. the number of training steps when different orthogonal regularization ratios are applied. We observe that the orthogonal regularization improves the overall performance. We also observe that, when applied relatively larger orthogonal regularization, the overall performance is about same but larger orthogonal regularization can still improve the converge speed of the model. It indicates that encouraging the orthogonality of latent column spaces effectively improves the quality of dynamic parameter generation/weighting.

\section{Conclusion}
Dynamic parameterization has become a popular way to leverage the deep neural networks' representation capacity. We design our network: DynInt, with two schemes to enhance model capacity and model dynamic feature interactions. With the proposed computational paradigm based on low-rank approximation, we reduce the memory cost in implementations. Our experimental results have demonstrated its superiority over
the state-of-art algorithms on real-world datasets, in terms of both model performance and efficiency.

In further work, We would like to study how to effectively and dynamically combine our model with other architectures to learn different types of interactions collectively. Beyond only modeling on fixed-length features, We also like to extend our architecture to learn on variable-length features for sequential behaviors in real-world problems.

%% \newpage
\bibliographystyle{ACM-Reference-Format}
\bibliography{dynint-ref.bib}
%% \nocite{*}

\end{document}